\documentclass[preprint,12pt, 3p,times]{elsarticle}

\usepackage{lineno,hyperref}
\modulolinenumbers[5]

\usepackage{amsmath}
\usepackage{makeidx}
\usepackage{graphics}
\usepackage{color}
\usepackage{listings}
\usepackage{bm}
\usepackage{url}
\usepackage{booktabs}
\usepackage{paralist}
\usepackage{amssymb}
\usepackage{amsthm} 
\usepackage{subcaption}
\usepackage{algorithm}
\usepackage{algorithmic}
\usepackage{lipsum}
\graphicspath{{fig/}}

\usepackage[percent]{overpic}

\usepackage{longtable}
\usepackage{dsfont}

\biboptions{sort&compress}
\newcommand*\diff{\mathop{}\!\mathrm{d}}

\begin{document}

\begin{frontmatter}

\title{ Probing bulk viscosity in relativistic flows }

\author{A. Gabbana }
\address{Universit\`a di Ferrara and INFN-Ferrara, I-44122 Ferrara,~Italy}
\address{Bergische Universit\"at Wuppertal, D-42119 Wuppertal,~Germany}
\author{D. Simeoni \fnref{corr-author} }
\address{Universit\`a di Ferrara and INFN-Ferrara, I-44122 Ferrara,~Italy}
\address{Bergische Universit\"at Wuppertal, D-42119 Wuppertal,~Germany}
\address{University of Cyprus, CY-1678 Nicosia,~Cyprus}
\author{S. Succi}
\address{Center for Life Nano Science @ La Sapienza, Italian Institute of Technology, Viale Regina Elena 295, I-00161 Roma,~Italy}
\address{Istituto Applicazioni del Calcolo, National Research Council of Italy, Via dei Taurini 19, I-00185 Roma,~Italy}
\author{R. Tripiccione}
\address{Universit\`a di Ferrara and INFN-Ferrara, I-44122 Ferrara,~Italy}
\fntext[corr-author]{Corresponding author. E-mail address: \href{mailto:d.simeoni@stimulate-ejd.eu}{d.simeoni@stimulate-ejd.eu}}
 
\begin{abstract}
We derive an analytical connection between kinetic relaxation rate and bulk viscosity of a 
relativistic fluid in $d$ spatial dimensions, all the way from the ultra-relativistic down to 
the near non-relativistic regime. Our derivation is based on both Chapman-Enskog asymptotic 
expansion and Grad's method of moments. We validate our theoretical results against a 
benchmark flow, providing further evidence of the correctness of the Chapman-Enskog approach; 
we define the range of validity of this approach and provide evidence of mounting departures 
at increasing Knudsen number. Finally, we present numerical simulations of transport processes 
in quark gluon plasmas, with special focus on the effects of bulk viscosity which might prove 
amenable to future experimental verification.
\end{abstract}

\end{frontmatter}

\newpage


\newpage

\section{Introduction}
In the last decade, relativistic hydrodynamics has received renewed attention and interest 
thanks to major breakthroughs in condensed matter, high-energy and gravitational physics \cite{romatschke-book-2019}. 
In particular, experimental data from the Relativistic Heavy-Ion Collider (RHIC) and the Large Hadron 
Collider (LHC), have triggered further developments in the study of viscous relativistic fluid dynamics, 
both at the level of theoretical formulations and for the development of robust numerical methods, to 
describe the collective behavior of quark gluon plasmas (QGP).

Several theoretical aspects related to a consistent formulation of 
dissipative relativistic hydrodynamics are still under debate in the literature \cite{muronga-prc-2007,
muronga-prc-2007b,betz-epj-2009,el-prc-2010, denicol-prl-2010, betz-epj-2011,denicol-prd-2012, molnar-prd-2014, 
tsumura-epj-2012,tsumura-prd-2015, jaiswal-prc-2013,jaiswal-prc-2013b,jaiswal-prc-2013c,bhalerao-prc-2013,
chattopadhyay-prc-2015, kikuchi-prc-2015, kikuchi-pla-2016}, including the correct derivation of the values 
of the transport coefficients as a function of the parameters defined at the level of kinetic theory 
\cite{mendoza-josm-2013, bhalerao-prc-2014, florkowski-prc-2015, perciante-josp-2017, ambrus-prc-2018b, 
gabbana-pre-2017b, gabbana-pre-2019, perciante-jop-2019, perciante-pa-2019}.

A proper understanding of transport properties is crucial for the study of the evolution and equilibration 
process of the quark gluon plasma produced in heavy ion collisions. The effects of shear viscosity 
on the elliptic flow parameters have been extensively studied by several authors \cite{heinz-prc-2006, 
romatschke-prl-2007, song-prc-2008, song-prc-2008b, molnar-jop-2008}. 
The relevance of bulk viscosity, mostly regarded as negligible in the earlier days, has attracted significant 
attention in recent years \cite{denicol-prc-2009,bozek-prc-2010,dobado-prd-2012,hostler-prc-2013,kadam-npa-2015}. 
For instance, it has been suggested that, near the critical point, 
bulk effects might be dominant over shear viscosity \cite{karsch-prb-2008, harvey-prl-2008}.
In this context, an accurate derivation of all transport coefficients and the availability of numerical tools 
capable of capturing the effects of bulk viscosity are desirable in a theoretical perspective and important 
also for the simulation of QGP.
As a side note, we remark that a complete analysis of the role of bulk viscosity in relativistic hydrodynamics 
could also be beneficial to the theoretical understanding of the accelerated expansion of the universe 
\cite{dou-aa-2011, gagnon-jcap-2011, velten-prd-2013, velten-ijgmmp-2014, atreya-jcap-2018}.

In this work, we perform the Chapman Enskog expansion to establish the analytic expression of bulk viscosity 
of a relativistic gas obeying an ideal equation of state and working in the single relaxation-time (SRT) approximation. 
The derivation is developed in a $(d+1)$ dimensional flat space time. 
While $d = 2, 3$ are the most relevant physical cases, it is nevertheless interesting from a theoretical point 
of view to consider the general $d$-dimensional case. 
Indeed, the dependence of bulk viscosity on the relativistic parameter $\zeta = \frac{m c^2}{k_B T}$  
(defined as the ratio between the particles rest energy and the thermal energy), is found to 
strongly depend on the dimensionality of the system.

Our analytical results are then compared and validated against numerical simulations, performed using a 
recently developed relativistic lattice kinetic scheme. We consider first a simple synthetic flow that 
we would like to suggest as a benchmark for the measurement and calibration of bulk viscosity and then 
a more complex flow with several features typical of QGP flows. This paper builds on previous work presented 
in \cite{gabbana-arxiv-2019}, enriched with an extended set of new numerical results. 

This paper is organized as follows: in section 2, we briefly summarize the procedure followed to derive 
the analytic form of bulk viscosity working in the single relaxation time approximation. In section 3, we
present a numerical validation of the analytical results, also providing an 
example of application for which these results are relevant and of practical interest.
Finally, conclusions and future developments are summarized in section 4.

\section{Relativistic Boltzmann Equation and Chapman Enskog expansion} \label{sec:hydro}

We consider a $(d+1)$ dimensional flat spacetime, in which a statistical description of a relativistic fluid
is given in terms of the particle distribution function $f( (x^{\alpha}), (p^{\alpha}) )$, 
depending on coordinates $\left( x^{\alpha} \right) = \left( c t, \bm{x} \right)$, with $c$ the light speed, 
and momenta $\left( p^{\alpha} \right) = \left( p^0, \bm{p} \right)$, with $\bm{x}, \bm{p} \in \mathbb{R}^d$.

The time evolution of the system is governed by the relativistic Boltzmann equation,
that we take in the Anderson-Witting \cite{anderson-witting-ph-1974a,anderson-witting-ph-1974b} 
SRT approximation:
\begin{equation}\label{eq:rta}
  p^{\alpha} \frac{\partial f}{\partial x^{\alpha}} 
  = 
  - \frac{p^{\mu} U_{\mu}}{\tau c^2} \left( f - f^{\rm eq} \right) \quad ,
\end{equation}
with $\tau$ the relaxation (proper-) time, $U^{\mu}$ the macroscopic fluid velocity, and $f^{\rm eq}$ 
the equilibrium distribution function, for which we take the normalized Maxwell-J\"uttner distribution:
\begin{equation}\label{eq:feq}
  f^{\rm eq} = \left(\frac{c}{k_{\rm B} T}\right)^d 
               \frac{n}{2^{\frac{d+1}{2}} \pi ^{\frac{d-1}{2}} \zeta ^{\frac{d+1}{2}} K_{\frac{d+1}{2}}(\zeta )} 
               \exp{ \left( - \frac{U^{\alpha} p_{\alpha}}{k_{\rm B} T} \right) } \quad .
\end{equation}
In the above $n$ is the particle number density, $T$ the temperature, $\zeta$ (already referred to in the Introduction) is 
the ratio between the rest mass of the particles $m$ and the temperature ( $\zeta = m c^2 / k_{\rm B} T$ ), $K_i(x)$ the 
modified Bessel function of the second kind of order $i$, and $k_{\rm B}$ the Boltzmann constant. 

The Anderson-Witting collisional operator ensures the local conservation of particle number, energy and momentum.
Dissipative effects are described by the energy momentum tensor, which in the Landau-Lifshitz frame admits the
following decomposition:
\begin{equation}\label{eq:energy-momentum-tensor}
  T^{\alpha \beta} 
  = 
  c \int f p^{\alpha} p^{\beta} \frac{\diff^d p}{p_0} 
  = 
  \frac{\epsilon}{c^2} U^{\alpha} U^{\beta} - \left( P + \varpi \right) \Delta^{\alpha \beta} + \pi^{< \alpha \beta >} \quad ,
\end{equation}
with $\epsilon$ the energy density, $P$ the hydrostatic pressure, and $\Delta^{\alpha\beta}$ 
the Minkowski-orthogonal projector with respect to the fluid velocity $U^{\alpha}$:
\begin{equation}
  \Delta^{\alpha\beta} = \eta^{\alpha\beta} - \frac{1}{c^2} U^\alpha U^\beta \quad ;
\end{equation}
$\eta^{\alpha \beta}$ is the metric tensor, that we define as
$\eta^{\alpha \beta} = \rm{diag}(1, -\mathds{1})$, $\mathds{1} = \left( 1, \dots, 1 \right) \in \mathbb{N}^d $. 
Finally, and most importantly in this treatment, the pressure deviator $\pi^{<\alpha \beta>}$ (here the $<..>$ 
parentheses represent the traceless symmetric contribution to $T^{\alpha \beta}$) and dynamic pressure $\varpi$ 
represent the non-equilibrium contribution to the energy momentum tensor, proportional to shear viscosity $\eta$ 
and bulk viscosity $\mu$, respectively. It can be shown \cite{cercignani-book-2002} that bulk viscosity connects 
dynamic pressure and the divergence of the velocity via the relation:
\begin{equation}\label{eq:dynamic-pressure}
  \varpi = - \mu \nabla^{\alpha\beta} \partial_\beta U_{\alpha} \quad .
\end{equation}

Asymptotic expansions are generally employed in order to establish a link between macroscopic equations and
the kinetic description. In the following, we perform the Chapman-Enskog expansion \cite{chapman-book-1970} to
determine an analytic expression putting in relation the bulk viscosity with the kinetic relaxation time parameter.
Here we confine ourselves to a summary of the main conceptual steps, while full details on the analytic procedure 
can be found in \cite{gabbana-arxiv-2019}.

The starting point is an expansion of the particle distribution function $f$ around equilibrium 
\begin{equation}\label{eq:f-approx-ce-main}
  f \sim f^{\rm eq} ( 1 + \phi ) \quad ,
\end{equation}
with $\phi$ of the order of the Knudsen number $\rm Kn$, defined as the ratio between the mean free path and a 
typical macroscopic lenght scale. Next, we plug Eq.~\ref{eq:f-approx-ce-main} into Eq.~\ref{eq:rta}, and retain 
only terms $\mathcal{O}(\rm{Kn})$:
\begin{equation}\label{eq:rta-rbe-ce}
  p^{\alpha} \frac{\partial f^{\rm eq}} {\partial x^{\alpha}}
  = 
  - \frac{p^{\alpha}U_{\alpha}}{c^{2} \tau} f^{\rm eq} \phi \quad .
\end{equation}
By combining the above with Eq.~\ref{eq:feq} lengthy but straightforward calculations allow to 
derive an analytic expression for $\phi$:
\begin{equation}\label{eq:phi}
  \phi = -\frac{c^2\tau}{p^\mu U_\mu}p^\alpha
        \left[    \frac{\partial_\alpha n}{n} 
                + \left(1-G_d\right)\frac{\partial_\alpha T}{T}
                + p^{\beta}\frac{U_{\beta}\partial_{\alpha}T}{k_{\rm B}T^2}
                - \frac{p^{\beta}\partial_{\alpha}U_{\beta}}{k_{\rm B}T}
        \right] \quad , 
\end{equation}
where 
\begin{equation}
  G_{d} 
  = 
  \frac{\epsilon + P}{P} 
  = 
  \zeta \frac{ K_{\frac{d+3}{2}}(\zeta) }{ K_{\frac{d+1}{2}}(\zeta) }  \quad .
\end{equation}

At this point it is possible to use Eq.~\ref{eq:f-approx-ce-main} to compute the energy-momentum tensor $T^{\alpha \beta}$
through its integral definition. Moreover, from Eq.~\ref{eq:energy-momentum-tensor} one can single out the dynamic pressure
by applying the projector $\Delta_{\alpha\beta}$, giving:
\begin{align}\label{eq:varpi-from-tensor}
  \varpi = - P - \frac{1}{d}\Delta_{\alpha\beta}T^{\alpha \beta} \quad .
\end{align}

By comparing Eq.~\ref{eq:dynamic-pressure} with Eq.~\ref{eq:varpi-from-tensor}, and matching term by term 
we identify the following analytic expression for bulk viscosity:
\begin{equation}\label{eq:mu-ce}
  \mu = P \tau \left[\frac{G_d-\zeta^2 K}{d} - \frac{ \zeta^2 -G_{d}^2 +(d+2) G_{d}}{ \zeta^2 -G_{d}^2 +(d+2) G_{d}-1} \right] 
  \quad \text{with} \quad 
  K = \int p^i p^i \frac{f^{\rm eq}}{p^\mu U_\mu} \frac{\diff^d p}{p_0} \quad.
\end{equation}
Following a similar procedure it is possible to extract a value also for the shear viscosity $\eta$, that we show here for completeness 
(further details on the derivation can be found in \cite{gabbana-arxiv-2019}):
\begin{equation}\label{eq:shear-ce}
  \eta = P \tau \left[\frac{G_d-\zeta^2 K}{d + 2}\right] 
\end{equation}
Grad's method of moments \cite{grad-cpam-1949} provides an alternate procedure to connect kinetic parameters with hydrodynamics coefficients.
Although there is a growing consensus on CE providing more accurate results with respect to Grad's method \cite{plumari-prc-2012,
tsumura-epj-2012,jaiswal-prc-2013b,kikuchi-pla-2016,tsumura-prd-2015,gabbana-pre-2017b,coelho-cf-2018,ambrus-prc-2018b,gabbana-arxiv-2019}
it is nevertheless interesting to compare the two. Following the procedure described in \cite{cercignani-book-2002}
we obtain the following expressions for the bulk and shear viscosity of a relativistic fluid in $(d+1)$ dimensions:
\begin{align}\label{eq:mu-shear-grad}
  \mu     &= P \tau \frac{\left(\zeta^2 (d-2 G_{d})+G_{d} (-d+G_{d}-1) (-d+2 G_{d}-2)\right)^2}{d \left(G_{d} (d-G_{d}+2)+\zeta^2-1\right)} \times \\
          &  \frac{1}{G_{d}^2 \left(d^2+8 d-2 \zeta^2+12\right)-G_{d} \left(d^2+d \left(5-3 \zeta^2\right)-10 \zeta^2+6\right)
                                    + \zeta^2 \left(-d+2 \zeta^2-2\right)-(d+6) G_{d}^3} \quad . \notag \\
  \eta    &= P \tau \frac{G_d^2}{(d+3) G_d + \zeta^2}
\end{align}
%
%
\begin{figure}[t]
  \centering
  \begin{overpic}[width=.99\columnwidth]{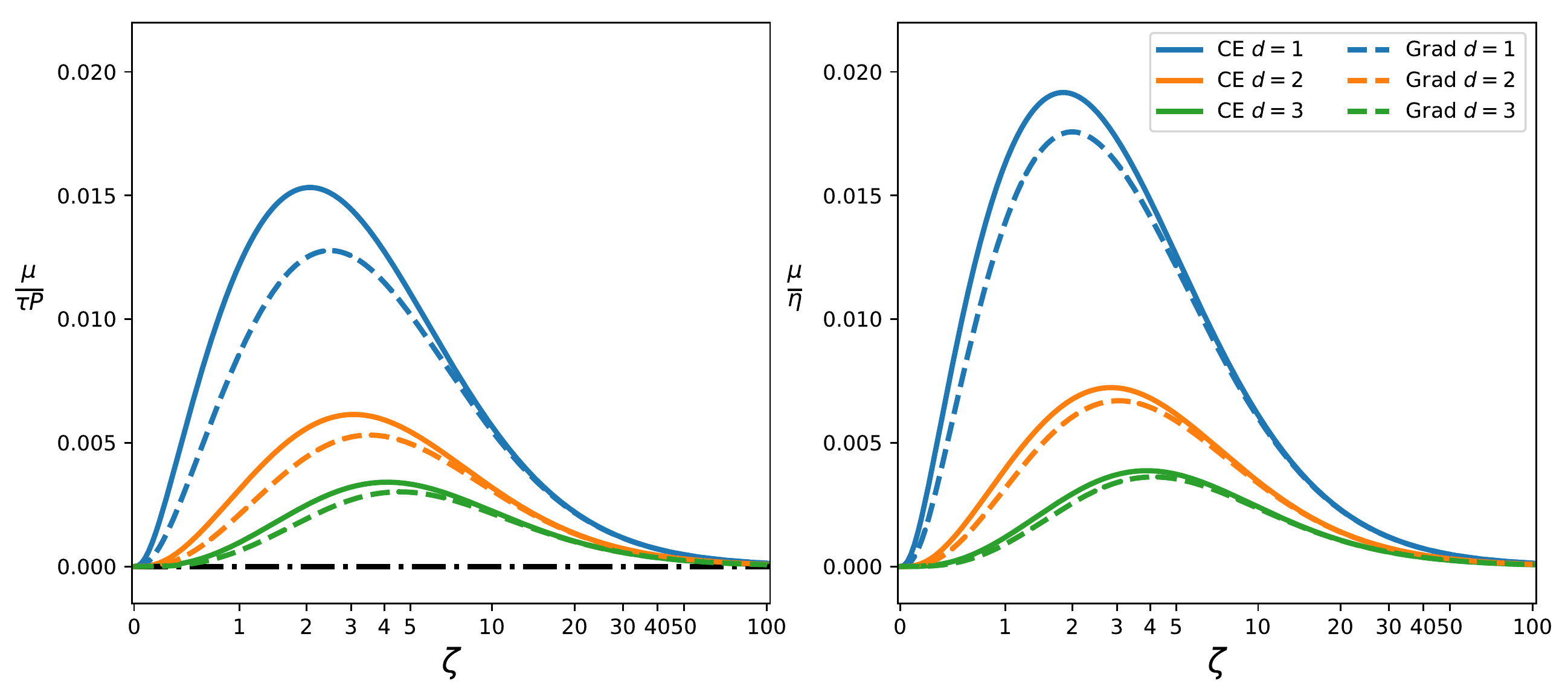}
        \put(1 ,41){(a)}
        \put(50,41){(b)}
  \end{overpic}  
   \caption{ On the left, bulk viscosity is plotted against the parameter $\zeta$ with a dependence given 
            by the CE analysis (thick lines) and Grad's method (dashed lines). The discrepancy between the two
            asymptotic expansions in more prominent in the mildly relativistic regime. The black dashed line represent 
            the viscosity in the limit of infinite spatial dimensions, common to both CE and Grad's method, 
            where the transport coefficient is independent of $\zeta$. 
            On the right, the ratio between the bulk and the shear viscosity ($\eta$) is plotted as a function of $\zeta$. 
            } \label{fig:1}          
\end{figure}
%
%

We compare the behavior of $\mu$ obtained using CE and Grad's method of moments
in $1, 2$ and $3$ space dimensions in Fig~\ref{fig:1}(a) . Both methods correctly reproduce the expected 
limiting behavior for which bulk viscosity vanishes in the ultra-relativistic
($\zeta \rightarrow 0$) and non-relativistic ($\zeta \rightarrow \infty$) limit. However there is an
intermediate region for which a non-zero bulk viscosity is predicted and for which the two derivations yield
different values for both the amplitude and the location of the peak.
Despite the bulky analytical expressions, the position of the maximum, $\zeta_{\rm max}$, 
is found to have a very simple linear dependence on the dimension of the system: 
$\zeta_{\rm max} = \alpha_1 d + \alpha_0$, with $\alpha_1$ consistent with $1$ in both cases, and 
$\alpha_0 \approx 0.744 $  for CE, $\alpha_0 \approx 1.235 $ for Grad. 
As an amusing theoretical remark, we also observe that, in the limit of infinite spatial dimensions, 
bulk viscosity vanishes for all values of $\zeta$.

Finally, we conclude our analysis pointing out one important limitation of the Anderson-Witting
collisional operator: since this model depends on one single mesoscopic parameter it follows that 
the relaxation rate will be the same for all the transport coefficients. As a consequence one cannot
tune independently shear and bulk viscosity; their ratio is shown in Fig.~\ref{fig:1}(b).

\section{Numerical Simulations}

In this section, we make use of a recently developed lattice kinetic solver \cite{gabbana-pre-2017,gabbana-pre-2017b}
and present results of numerical tests which aim at i) cross-checking and validating the
analytical results presented in the previous section and ii) providing an example of a realistic application 
to the physics of quark gluon plasmas. We first give a short overview of the lattice kinetic algorithm that we have used 
(a detailed derivation can be found in \cite{gabbana-arxiv-2019}) and then proceed to present the numerical tests.

\subsection{I: Numerical Scheme}

The relativistic lattice Boltzmann method (RLBM) is a computationally efficient approach to 
dissipative relativistic hydrodynamics. It is based on a mesoscale approach, and therefore
it has the advantage, with respect to other relativistic hydrodynamic solvers, that 
the emergence of viscous effects does not break relativistic invariance and causality, 
since space and time are treated on the same footing, i.e. both via first order derivatives.

This numerical method solves a minimal version of Eq.~\ref{eq:rta}, in which the 
discretization of the microscopic momentum vector on a Cartesian grid is coupled with
a Gauss-type quadrature (see \cite{gabbana-pre-2017b} and \cite{gabbana-arxiv-2019} for
the formal analytic derivation) which ensures the preservation of the lower 
(hydrodynamics) moments of the particle distribution:
\begin{align}\label{eq:discrete_moment}
  f_i(\bm{x} + \bm{v}_i \Delta t, t+\Delta t) 
  =
  f_i(\bm{x}, t) - \Delta t\frac{p^\mu_i U^\mu}{c p_i^0 \tau} \left(f_i(\bm{x}, t) - f_i^{\rm eq}(\bm{x}, t) \right) \quad i = 1,2,\dots M \quad .
\end{align}
In the above $\bm{v}_{i} = \bm{p}_i / p^0_i$ are the microscopic velocities, chosen in such a
way to i) preserve exact streaming (meaning that (pseudo)-particles travel in one time step along 
constant streamlines $\bm{x} + \bm{v}_i \Delta t$ from a point of the grid to 
another point of the grid) ii) together with an appropriate set of weights $w_i$ reproduce 
correctly the moments of the particle distribution up to order $N$.
Given this two conditions, $f_i^{\rm eq}$ can be defined as the discrete version of 
a polynomial expansion of the equilibrium distribution:
\begin{align}\label{eq:feq_exp_trunc}
  f_i^{\rm eq} = w_i \sum_{k=1}^{N} a_{(k)}(U^\mu,T) J^{(k)}(p^\mu_i) \quad ;
\end{align}
refer to Appendix~F and ~G in \cite{gabbana-arxiv-2019} for the definition of the 
polynomials and the projection coefficients used in the expansion.
The numerical analysis presented in the coming section is based on numerical
simulations making use of third order quadratures ($N = 3$), which are listed in 
Appendix~H in \cite{gabbana-arxiv-2019}.

The time evolution of Eq.~\ref{eq:discrete_moment} follows the collide-streaming 
paradigm typical of classic Lattice Boltzmann schemes. At each time step, and for
each grid cell, we need to compute the macroscopic fields associated to the particle distribution.
In order to do so we start by computing the first and second moment of distribution:
$$ 
  N^{\alpha} = \sum_i f_i p^{\alpha}_i \quad , \quad T^{\alpha \beta} = \sum_i f_i p^{\alpha}_i p^{\beta}_i \quad .
$$
From the definition of the energy-momentum tensor in the Landau frame (Eq.~\ref{eq:energy-momentum-tensor})
we compute the energy density $\epsilon$ and the four velocity $U^{\alpha}$ by solving the eigenvalue problem
$$ 
  \epsilon U^{\alpha} = T^{\alpha \beta} U_{\beta} \quad ,
$$
where $\epsilon$ corresponds to the largest eigenvalue of $T^{\alpha \beta}$.
The particle density $n$ comes from the definition of the first order moment, while temperature 
and pressure follow from the ideal equation of state:
\begin{align}\label{eq:eqstate}
  P        &= n k_B T      \\
  \epsilon &= P(G_d - 1) \quad \notag. 
\end{align}
At this stage it is possible to compute the polynomial expansion of the equilibrium distribution,
defined in Eq.~\ref{eq:feq_exp_trunc}, and use it to evolve Eq.~\ref{eq:discrete_moment}.

\subsection{II: Validation and Calibration}

Following the same approach used in previous works on the analysis of shear viscosity and
thermal conductivity \cite{gabbana-pre-2017b,coelho-cf-2018,gabbana-pre-2019}, we compare our analytical 
predictions for bulk viscosity with data from numerical simulations.

We consider a simple synthetic flow describing a time-decaying sinusoidal wave in a 
$d$ dimensional periodic domain; this flow is characterized by sizeable velocity gradients, allowing
the detection of physical effects due to a non zero bulk viscosity.  
The initial conditions for the benchmark are as follows:
\begin{align}
  u_x &= v_0 \sin\left(\frac{2\pi}{L}x\right)       \quad x \in [0, L] \quad , \\
  u_i &= 0 \quad \forall i \neq x \quad , \notag
\end{align}
with $v_0$ a given initial velocity, and with constant initial values for both particle density and temperature.
%
%
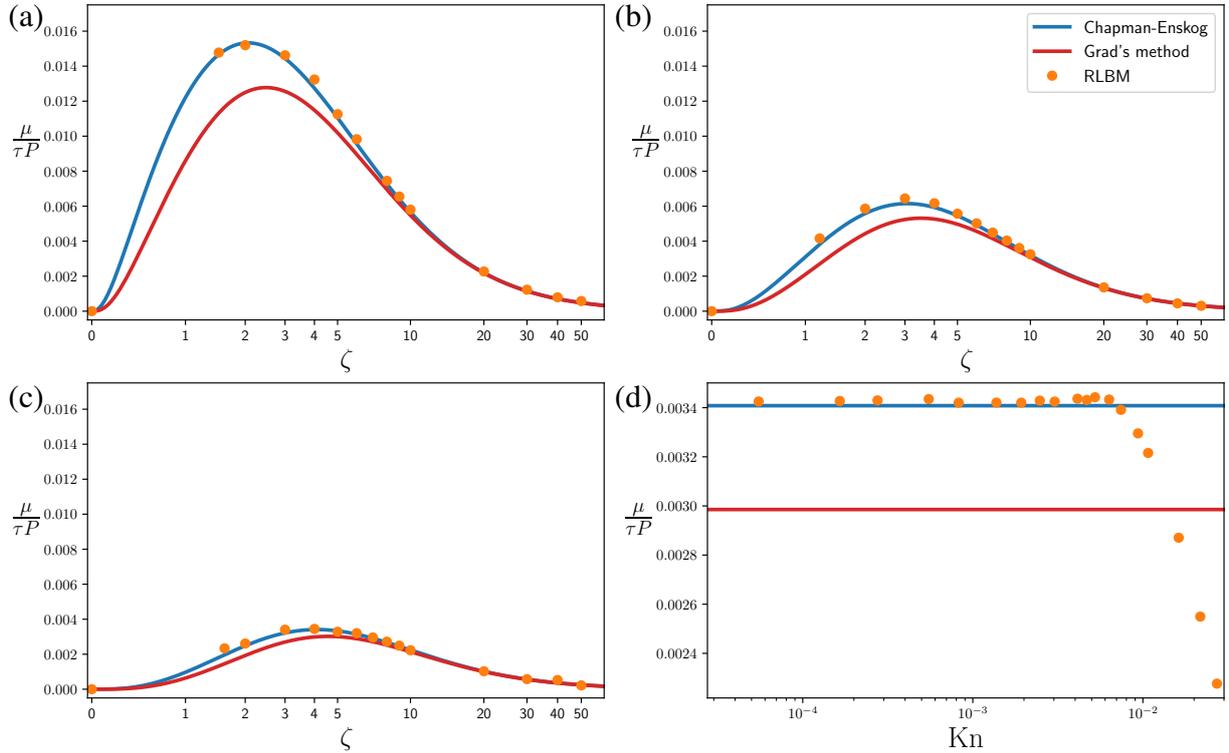
\begin{figure}[htb]
\centering
  \begin{overpic}[width=.99\columnwidth]{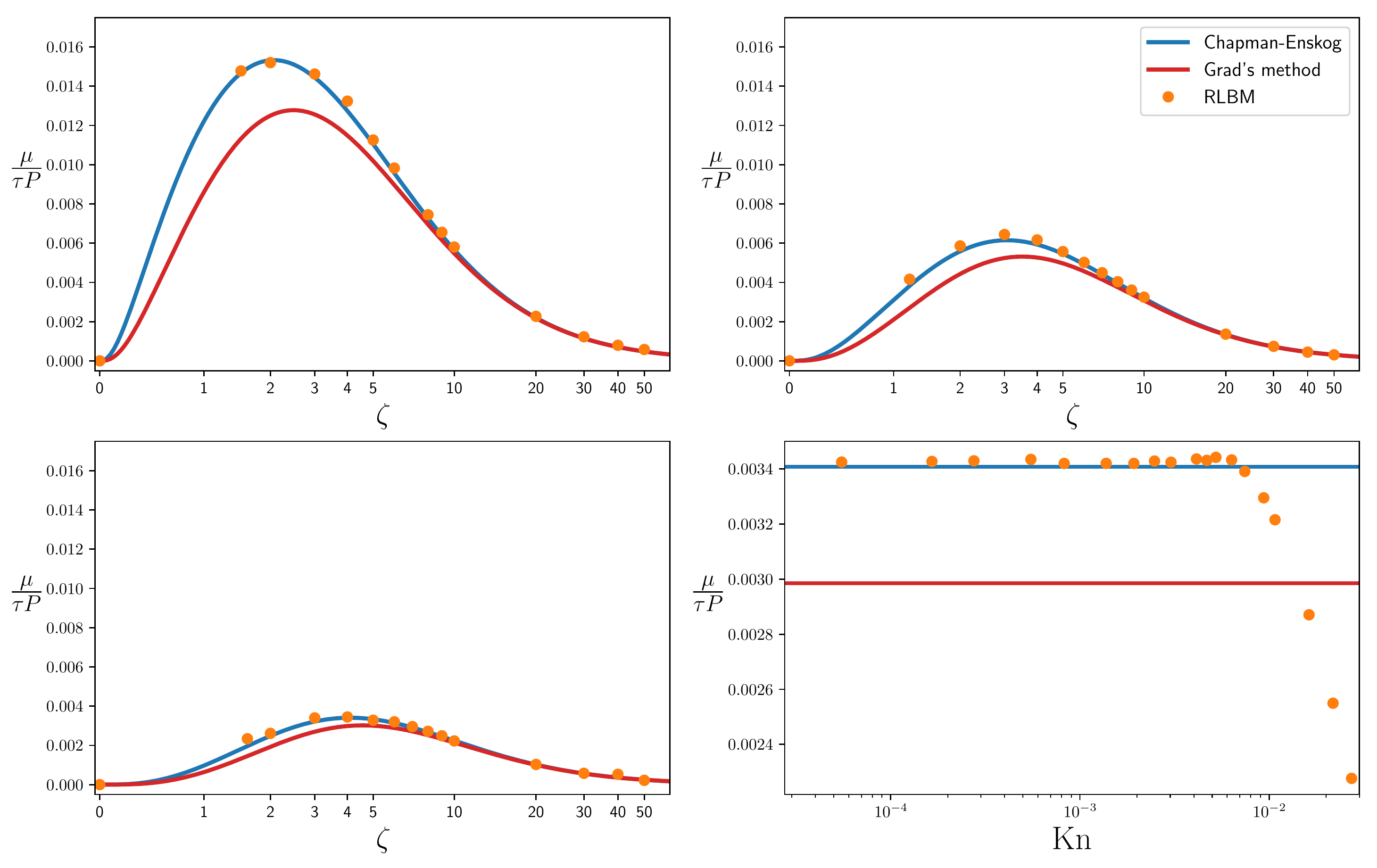}
        \put( 0.5, 60){(a)}
        \put(49.5, 60){(b)}
        \put( 0.5, 29){(c)}
        \put(49.5, 29){(d)}
  \end{overpic}  
\caption{ Numerical estimate of the (non-dimensional) bulk viscosity for a relativistic gas
          in (1 + 1), (2 + 1)  and (3 + 1) dimensions, shown respectively in panel (a), (b) and (c).
          The results are in agreement with CE analysis. In panel (d) we consider the specific case $\zeta = 4$
          in three dimensions and show how the estimate for $\mu$ varies as a function of the Knudsen number.
          We estimate the Knudsen number using $\rm{Kn} = <v> \tau / L$, where $<v>$ is an estimate of the 
          mean velocity of the particles of the relativistic fluid \cite{cercignani-book-2002}, and $\rm L$ 
          the typical lenght scale of the system, for which we consider the wavelength of the sine wave. 
          One can see that as $\rm{Kn}$ increases the first order approximation given by both CE and Grad 
          is no longer valid. 
        } \label{fig:2}
\end{figure}
%

In order to numerically evaluate the dynamic pressure, we introduce the definition of the energy-momentum
tensor at the equilibrium $T_{\rm E}^{\alpha \beta}$, which follows from Eq.~\ref{eq:energy-momentum-tensor}:
\begin{equation}\label{eq:energy-momentum-tensor-eq}
  T^{\alpha \beta}_{\rm E}
  =
  c \int f^{\rm eq} p^{\alpha} p^{\beta} \frac{\diff^d p}{p_0} 
  =
  \frac{\epsilon}{c^2} U^{\alpha} U^{\beta} - P \Delta^{\alpha \beta}  \quad .
\end{equation}
The dynamic pressure can then be expressed as the trace of the difference between the energy momentum 
tensor and its equilibrium counterpart:
\begin{equation}
  \varpi = -\frac{1}{d}( T^{\mu}_{\mu} - {T_{\rm E}}_{\mu}^{\mu})  \quad .
\end{equation}

When considering flows at sufficiently low speeds ($v_0 << c$), it is reasonable to approximate the 
relativistic divergence $\nabla^{\alpha\beta} \partial_\beta U_{\alpha}$ with its non-relativistic counterpart. 
It follows that we can numerically measure $\nabla^{\alpha\beta} \partial_\beta U_{\alpha}$ to good accuracy at 
each time step of the simulations, thus allowing an estimate of $\mu$ directly from Eq.~\ref{eq:dynamic-pressure}:
\begin{align}
  \mu = -\frac{\varpi}{\nabla^{\alpha\beta} \partial_\beta U_{\alpha}} \quad .
\end{align}

We have performed several simulations varying the mesoscopic parameters $\tau$ and $\zeta$ and extracted
the expression for $\mu$ as a function of $\zeta$ in various spatial dimensions. 
Our results, see Fig.~\ref{fig:2}, confirm that the CE analysis is in excellent agreement with numerical results. 

We point out that the choice of the relaxation time $\tau$ is key to obtain accurate results. The linear 
relationship between $\mu$ and $\tau$ holds as long as the assumptions 
made in the Chapman Enskog analysis remain valid, in particular the assumption of 
small Knudsen numbers. Conversely, for large values of $\tau$, that is for regimes where a purely
hydrodynamic treatment becomes questionable, the relation between the transport coefficients and the 
relaxation time is expected to depart from linearity \cite{ambrus-prc-2018b}.
This behavior is measured in Fig.~\ref{fig:2}(d), for a specific case in three-dimensions at $\zeta = 4$:
we plot the fitted value for $\mu / \tau P$ against the Knudsen number $\rm Kn$, clearly showing that for $\rm{Kn} \gtrapprox 0.01$
numerical data start to diverge from the CE prediction. One can expect better agreement when including higher 
order terms in the ansatz in Eq.~\ref{eq:f-approx-ce-main}, although this topic is still under debate in the 
literature since gradient-expansions are notoriously divergent \cite{denicol-arxiv-2016,noronha-npa-2017}. 

\subsection{III: Application}

%
\begin{figure}[htb!]
  \centering
  \hspace*{0.35cm}
  \includegraphics[width=0.99\textwidth]{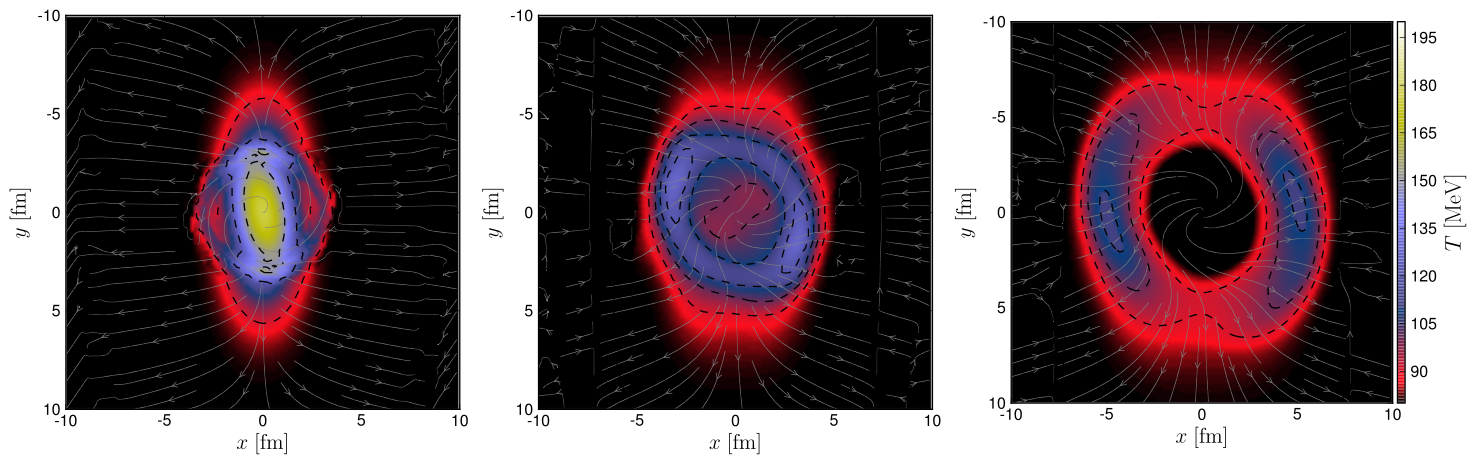}
  \includegraphics[width=0.99\textwidth]{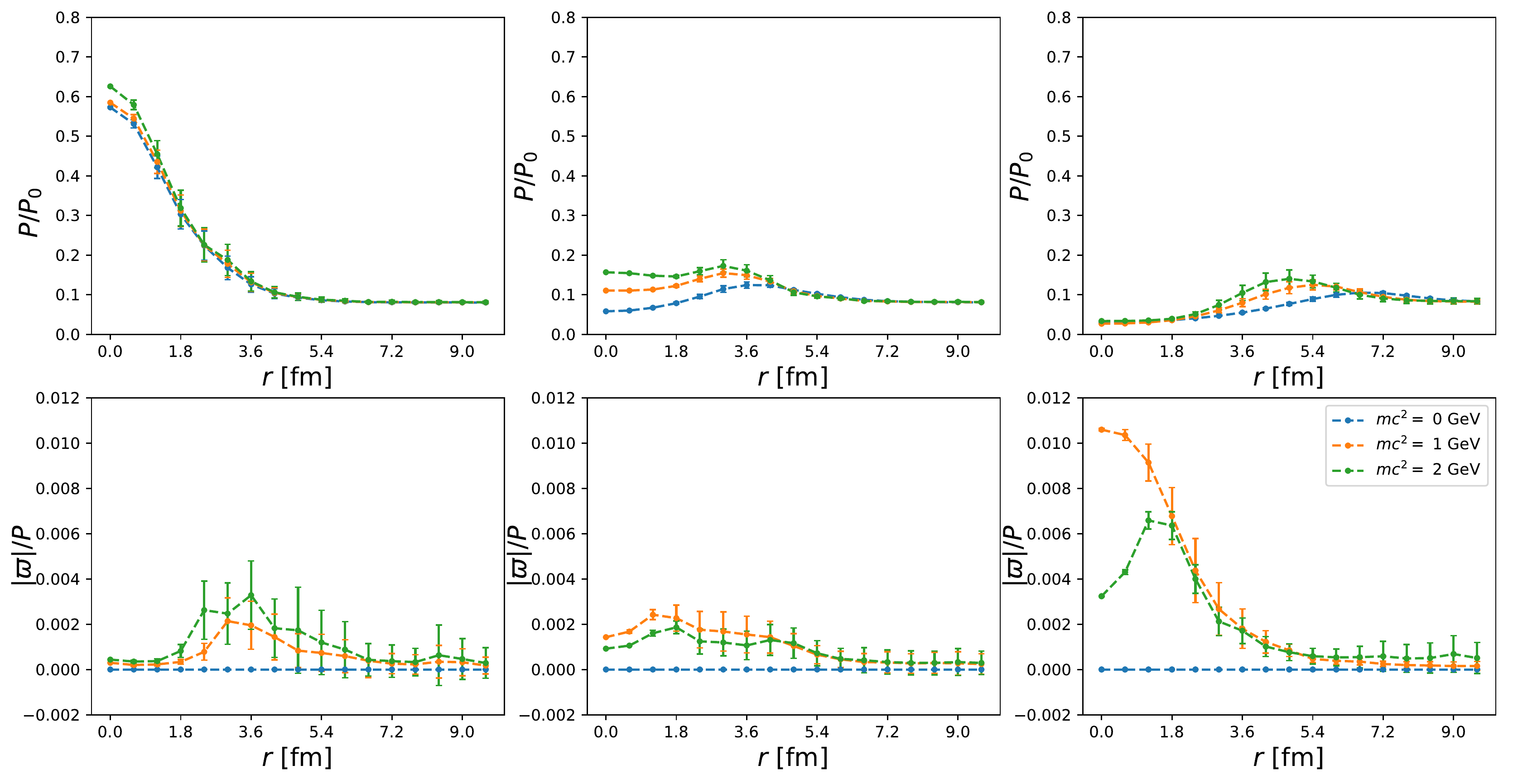}
  \caption{ Dynamic evolution of a fireball of QGP at three different time steps: $t = 1~\rm{fm/c}$ on 
            the left panels, $t = 4~\rm{fm/c}$ on the central panels, $t = 8~\rm{fm/c}$ on the right panels.
            In the top plots is shown the evolution of both the temperature (color map) and the velocity 
            (gray arrows) fields, on a slice of the system. The two bottom lines of plots show the time 
            evolution of respectively the hydrostatic pressure (normalized respect to the initial pressure 
            value at center) and ratio of dynamic to hydrostatic pressure, for three different values of 
            the particle mass $m$. While for $m=0$ the dynamic pressure is always zero, it is not the case 
            when $m \neq 0$. This shows that in this kind of dynamic the bulk viscosity plays a subtle, but 
            still relevant role.  
    }\label{fig:3}  
\end{figure} 
%

In this section, we present a second simulation example where we consider a qualitative description 
of a relativistic elliptic flow, thus mimicking the evolution of the initial stages of heavy-ion collisions.

We adopt the same numerical setup used in a series of studies on spin-polarized relativistic flows 
\cite{florkowski-prc-2018,florkowski-prd-2018,florkowski-arxiv-2019}, neglecting however quantum effects.
The equation of state used in these simulation is the ideal one showed in Eq~\ref{eq:eqstate}. Simulations 
of elliptic flows with different equation of state implementations can be found for example 
in \cite{niemi-prl-2011}. 

The initial conditions are given by a Gaussian distribution for both the temperature and density profiles
\begin{equation}
  T = T_{\rm 0}~g(x,y,z), 
  \quad 
  n = n_{\rm 0}~g(x,y,z), 
  \quad 
  g(x,y,z) = \exp{(-\frac{x^2}{2 \sigma_x^2} -\frac{y^2}{2 \sigma_y^2} -\frac{z^2}{2 \sigma_z^2})} \quad ,
\end{equation}
with $\sigma_x = 1~\rm{fm}$, $\sigma_y = 2.6~\rm{fm}$ and $\sigma_z = 2~\rm{fm}$.
The resulting ellipsoid represents the overlapping zone in the collision between two heavy nuclei,
with the fluid representing the product of such a collision, a hot, dense and strongly anisotropic ``fireball'' of QGP.
The initial temperature at the origin of the axis is $T_0 = 200~{\rm MeV}$, with
$n_0 = 4 \cdot 10^{-3} ~{\rm fm^{-3}}$. In order to avoid numerical instabilities we include
a background temperature of $T = 80~{\rm MeV}$ and density of $n = 1 \cdot 10^{-3}~{\rm fm^{-3}}$.

The initial velocity profile is given by:
\begin{equation}
  U^{\alpha} = \gamma \left( 1, - \Omega(r) y, \Omega(r) x, 0 \right)  ,
  \quad 
  \Omega(r) = \frac{1}{r} \tanh{(\frac{r}{r_0})} \quad .
\end{equation}
with $r = \sqrt{x^2 + y^2}$ the distance from the center of the vortex in the transverse plane, 
and $r_0$ a parameter controlling the strength of the flow.
This initial condition, due to the limit posed by the speed of light,
is physically meaningful only inside a  maximum radius $R < 1 / \Omega$; 
in what follows we use $R = 3~{\rm fm}$ and $r_0 = 6~{\rm fm}$.
We consider a viscous regime where the ratio between the shear viscosity $\eta$ and the 
entropy density $s$ is kept fixed at $2 / (4 \pi)$, comparing simulations for different
values of the rest mass of the particles in the fluid, with respectively $mc^2 = 0, 1, 2~\rm{GeV}$.
All simulations pertain to a cubic domain of size $20~{\rm fm}$, using $128^3$ grid points.

The top-panel in Fig.~\ref{fig:3} shows the temperature profile of the system in the $x-y$ plane at $z = 0~{\rm fm}$
at three different time steps, from left to right $t = 1,~4,~8~\rm{fm/c}$. 
In the QGP framework one is interested in measuring the translation of the initial spatial anisotropy 
into a momentum space anisotropy (which can be measured in experiments).
The discretization of the momentum space, upon which our numerical scheme relies upon, does not
allow to perform direct measurements of the elliptic flow coefficients;
therefore here we will only show that bulk viscosity effects can indeed 
be detected and measured, leaving a more detailed analysis to future works.
With this aim, we perform measurements of macroscopic observables spatially averaged at 
fixed radial distance from the center of the ellipsoid. 
Fig.~\ref{fig:3}-center and lower panel- show respectively the hydrostatic pressure and the 
ratio between dynamic and hydrostatic pressure. Vertical bars, representing the variance, are larger at radial 
distances where the flow still exhibits a significant spatial anisotropy. 
The effect of the rest mass of the particles in the fluid is evident: the dynamic slows down when heavier particles are 
taken in consideration. For massless particles, the ratio between the dynamic pressure is zero up to numerical accuracy,
independently of the radial distance, as expected from the analytical and numerical analysis presented in the 
previous sections. On the other hand, for massive particles, non-equilibrium contributions due to bulk viscosity become noticeable. 
In these cases  $\varpi$ strongly varies across the domain, taking values larger than $1\%$ of the hydrostatic pressure.

We stress here that although the dissipative dynamic in the system is inherently connected to both shear and bulk effects (and due to the single 
relaxation time nature of the numerical scheme a separate tuning of the two effects is not possible) the specific effects of bulk viscosity are singled out by the behavior of the dynamic pressure $\varpi$ via Eq~\ref{eq:dynamic-pressure}, since this quantity is affected only by the value of $\mu$.

\section{Conclusions}
Summarizing, in this work we have highlighted the role of bulk viscosity $\mu$ on the dynamics of a relativistic monoatomic gas. Using the Chapman Enskog expansion and Grad's method of moments, we have presented an analytical formulation in (d+1) dimensions, showing the dependence of bulk viscosity on the kinetic relaxation time $\tau$ and  the relativistic parameter $\zeta = m c^2 / k_B T $. 

Our analysis shows that, at variance with both the ultra-relativistic and non-relativistic regimes, there is a region in $\zeta$ space where bulk viscosity is non zero, whose location and extension depend on the dimensionality of the system. 
Next, in order to discern between the two expansion methods, a numerical validation has been presented. Once more, the correctness of the Chapman Enskog analysis over Grad's method has been proved, in analogy with what happens for shear viscosity and thermal conductivity; this result paves the way to the correct reproduction of viscous effects in relativistic simulations. In the same context, the measure of $\mu$ at different values of $\tau$ has allowed testing the first order approximations in both CE and Grad's theory, clearly identifying the kinematic range where a hydrodynamic description is appropriate. Finally, a more realistic benchmark has been presented in the framework of QGP physics. In detail, a strongly anisotropic hot dense plasma, resulting from the Lorentz contraction of heavy-ion collisions, has been simulated, highlighting the presence of bulk-related viscous effects on the transport properties of the QGP. 

One limit of SRT models is that they link multiple hydrodynamic coefficients to a single relaxation time, 
thus preventing the independent tuning of two viscosities. For this reason, the development 
of a multi-relaxation time (MRT) numerical scheme, and the corresponding derivation of
transport coefficients, is highly desirable for future studies of relativistic transport phenomena.

\section*{Acknowledgment}

The authors would like to thank Victor Ambru\c{s} for useful discussions.
DS has been supported by the European Union's Horizon 2020 research and
innovation programme under the Marie Sklodowska-Curie grant agreement No. 765048.
SS acknowledges funding from the European Research Council under the European
Union's Horizon 2020 framework programme (No. P/2014-2020)/ERC Grant Agreement No. 739964 (COPMAT).
All the numerical work has been performed on the COKA computing cluster at Universit\`a di Ferrara. 

\newpage
\section*{References}

\bibliography{biblio}

\begin{thebibliography}{10}
\expandafter\ifx\csname url\endcsname\relax
  \def\url#1{\texttt{#1}}\fi
\expandafter\ifx\csname urlprefix\endcsname\relax\def\urlprefix{URL }\fi
\expandafter\ifx\csname href\endcsname\relax
  \def\href#1#2{#2} \def\path#1{#1}\fi

\bibitem{romatschke-book-2019}
P.~Romatschke, U.~Romatschke, Relativistic Fluid Dynamics In and Out of
  Equilibrium: And Applications to Relativistic Nuclear Collisions, Cambridge
  University Press, 2019.
\newblock \href {http://dx.doi.org/10.1017/9781108651998}
  {\path{doi:10.1017/9781108651998}}.

\bibitem{muronga-prc-2007}
A.~Muronga, Relativistic dynamics of nonideal fluids: Viscous and
  heat-conducting fluids. i. general aspects and $3+1$ formulation for nuclear
  collisions, Phys. Rev. C 76 (2007) 014909.
\newblock \href {http://dx.doi.org/10.1103/PhysRevC.76.014909}
  {\path{doi:10.1103/PhysRevC.76.014909}}.

\bibitem{muronga-prc-2007b}
A.~Muronga, Relativistic dynamics of non-ideal fluids: Viscous and
  heat-conducting fluids. ii. transport properties and microscopic description
  of relativistic nuclear matter, Phys. Rev. C 76 (2007) 014910.
\newblock \href {http://dx.doi.org/10.1103/PhysRevC.76.014910}
  {\path{doi:10.1103/PhysRevC.76.014910}}.

\bibitem{betz-epj-2009}
B.~Betz, D.~Henkel, D.~H. Rischke, Complete second-order dissipative fluid
  dynamics, Journal of Physics G: Nuclear and Particle Physics 36~(6) (2009)
  064029.
\newblock \href {http://dx.doi.org/10.1088/0954-3899/36/6/064029}
  {\path{doi:10.1088/0954-3899/36/6/064029}}.

\bibitem{el-prc-2010}
A.~El, Z.~Xu, C.~Greiner, Extension of relativistic dissipative hydrodynamics
  to third order, Phys. Rev. C 81 (2010) 041901.
\newblock \href {http://dx.doi.org/10.1103/PhysRevC.81.041901}
  {\path{doi:10.1103/PhysRevC.81.041901}}.

\bibitem{denicol-prl-2010}
G.~S. Denicol, T.~Koide, D.~H. Rischke, Dissipative relativistic fluid
  dynamics: A new way to derive the equations of motion from kinetic theory,
  Phys. Rev. Lett. 105 (2010) 162501.
\newblock \href {http://dx.doi.org/10.1103/PhysRevLett.105.162501}
  {\path{doi:10.1103/PhysRevLett.105.162501}}.

\bibitem{betz-epj-2011}
{Betz, B.}, {Denicol, G.S.}, {Koide, T.}, {Moln\'ar, E.}, {Niemi, H.},
  {Rischke, D.H.}, Second order dissipative fluid dynamics from kinetic theory,
  EPJ Web of Conferences 13 (2011) 07005.
\newblock \href {http://dx.doi.org/10.1051/epjconf/20111307005}
  {\path{doi:10.1051/epjconf/20111307005}}.

\bibitem{denicol-prd-2012}
G.~S. Denicol, H.~Niemi, E.~Moln\'ar, D.~H. Rischke, Derivation of transient
  relativistic fluid dynamics from the boltzmann equation, Phys. Rev. D 85
  (2012) 114047.
\newblock \href {http://dx.doi.org/10.1103/PhysRevD.85.114047}
  {\path{doi:10.1103/PhysRevD.85.114047}}.

\bibitem{molnar-prd-2014}
E.~Moln\'ar, H.~Niemi, G.~S. Denicol, D.~H. Rischke, Relative importance of
  second-order terms in relativistic dissipative fluid dynamics, Phys. Rev. D
  89 (2014) 074010.
\newblock \href {http://dx.doi.org/10.1103/PhysRevD.89.074010}
  {\path{doi:10.1103/PhysRevD.89.074010}}.

\bibitem{tsumura-epj-2012}
K.~Tsumura, T.~Kunihiro, {Derivation of relativistic hydrodynamic equations
  consistent with relativistic Boltzmann equation by renormalization-group
  method}, The European Physical Journal A 48~(11) (2012) 162.
\newblock \href {http://dx.doi.org/10.1140/epja/i2012-12162-x}
  {\path{doi:10.1140/epja/i2012-12162-x}}.

\bibitem{tsumura-prd-2015}
K.~Tsumura, Y.~Kikuchi, T.~Kunihiro, Relativistic causal hydrodynamics derived
  from boltzmann equation: A novel reduction theoretical approach, Phys. Rev. D
  92 (2015) 085048.
\newblock \href {http://dx.doi.org/10.1103/PhysRevD.92.085048}
  {\path{doi:10.1103/PhysRevD.92.085048}}.

\bibitem{jaiswal-prc-2013}
A.~Jaiswal, R.~S. Bhalerao, S.~Pal, Complete relativistic second-order
  dissipative hydrodynamics from the entropy principle, Phys. Rev. C 87 (2013)
  021901.
\newblock \href {http://dx.doi.org/10.1103/PhysRevC.87.021901}
  {\path{doi:10.1103/PhysRevC.87.021901}}.

\bibitem{jaiswal-prc-2013b}
A.~Jaiswal, Relativistic dissipative hydrodynamics from kinetic theory with
  relaxation-time approximation, Phys. Rev. C 87 (2013) 051901.
\newblock \href {http://dx.doi.org/10.1103/PhysRevC.87.051901}
  {\path{doi:10.1103/PhysRevC.87.051901}}.

\bibitem{jaiswal-prc-2013c}
A.~Jaiswal, Relativistic third-order dissipative fluid dynamics from kinetic
  theory, Phys. Rev. C 88 (2013) 021903.
\newblock \href {http://dx.doi.org/10.1103/PhysRevC.88.021903}
  {\path{doi:10.1103/PhysRevC.88.021903}}.

\bibitem{bhalerao-prc-2013}
R.~S. Bhalerao, A.~Jaiswal, S.~Pal, V.~Sreekanth, Particle production in
  relativistic heavy-ion collisions: A consistent hydrodynamic approach, Phys.
  Rev. C 88 (2013) 044911.
\newblock \href {http://dx.doi.org/10.1103/PhysRevC.88.044911}
  {\path{doi:10.1103/PhysRevC.88.044911}}.

\bibitem{chattopadhyay-prc-2015}
C.~Chattopadhyay, A.~Jaiswal, S.~Pal, R.~Ryblewski, Relativistic third-order
  viscous corrections to the entropy four-current from kinetic theory, Phys.
  Rev. C 91 (2015) 024917.
\newblock \href {http://dx.doi.org/10.1103/PhysRevC.91.024917}
  {\path{doi:10.1103/PhysRevC.91.024917}}.

\bibitem{kikuchi-prc-2015}
Y.~Kikuchi, K.~Tsumura, T.~Kunihiro, Derivation of second-order relativistic
  hydrodynamics for reactive multicomponent systems, Phys. Rev. C 92 (2015)
  064909.
\newblock \href {http://dx.doi.org/10.1103/PhysRevC.92.064909}
  {\path{doi:10.1103/PhysRevC.92.064909}}.

\bibitem{kikuchi-pla-2016}
Y.~Kikuchi, K.~Tsumura, T.~Kunihiro, Mesoscopic dynamics of fermionic cold
  atoms — quantitative analysis of transport coefficients and relaxation
  times, Physics Letters A 380~(24) (2016) 2075 -- 2080.
\newblock \href {http://dx.doi.org/10.1016/j.physleta.2016.04.027}
  {\path{doi:10.1016/j.physleta.2016.04.027}}.

\bibitem{mendoza-josm-2013}
M.~Mendoza, I.~Karlin, S.~Succi, H.~J. Herrmann, Ultrarelativistic transport
  coefficients in two dimensions, Journal of Statistical Mechanics: Theory and
  Experiment 2013 (2013) P02036.
\newblock \href {http://dx.doi.org/10.1088/1742-5468/2013/02/p02036}
  {\path{doi:10.1088/1742-5468/2013/02/p02036}}.

\bibitem{bhalerao-prc-2014}
R.~S. Bhalerao, A.~Jaiswal, S.~Pal, V.~Sreekanth, Relativistic viscous
  hydrodynamics for heavy-ion collisions: A comparison between the
  chapman-enskog and grad methods, Phys. Rev. C 89 (2014) 054903.
\newblock \href {http://dx.doi.org/10.1103/PhysRevC.89.054903}
  {\path{doi:10.1103/PhysRevC.89.054903}}.

\bibitem{florkowski-prc-2015}
W.~Florkowski, A.~Jaiswal, E.~Maksymiuk, R.~Ryblewski, M.~Strickland,
  Relativistic quantum transport coefficients for second-order viscous
  hydrodynamics, Phys. Rev. C 91 (2015) 054907.
\newblock \href {http://dx.doi.org/10.1103/PhysRevC.91.054907}
  {\path{doi:10.1103/PhysRevC.91.054907}}.

\bibitem{perciante-josp-2017}
A.~L. Garc{\'i}a-Perciante, A.~R. M{\'e}ndez, E.~Escobar-Aguilar, Heat flux for
  a relativistic dilute bidimensional gas, Journal of Statistical Physics 167
  (2017) 123--134.
\newblock \href {http://dx.doi.org/10.1007/s10955-017-1742-x}
  {\path{doi:10.1007/s10955-017-1742-x}}.

\bibitem{ambrus-prc-2018b}
V.~E. Ambru\c{s}, Transport coefficients in ultrarelativistic kinetic theory,
  Phys. Rev. C 97 (2018) 024914.
\newblock \href {http://dx.doi.org/10.1103/PhysRevC.97.024914}
  {\path{doi:10.1103/PhysRevC.97.024914}}.

\bibitem{gabbana-pre-2017b}
A.~Gabbana, M.~Mendoza, S.~Succi, R.~Tripiccione, Kinetic approach to
  relativistic dissipation, Phys. Rev. E 96 (2017) 023305.
\newblock \href {http://dx.doi.org/10.1103/PhysRevE.96.023305}
  {\path{doi:10.1103/PhysRevE.96.023305}}.

\bibitem{gabbana-pre-2019}
A.~Gabbana, D.~Simeoni, S.~Succi, R.~Tripiccione, Relativistic dissipation
  obeys chapman-enskog asymptotics: Analytical and numerical evidence as a
  basis for accurate kinetic simulations, Phys. Rev. E 99 (2019) 052126.
\newblock \href {http://dx.doi.org/10.1103/PhysRevE.99.052126}
  {\path{doi:10.1103/PhysRevE.99.052126}}.

\bibitem{perciante-jop-2019}
A.~L. Garc{\'{\i}}a-Perciante, L.~Franco-P{\'{e}}rez, A.~R. M{\'{e}}ndez, Bulk
  viscosity in 2d relativistic uids: the effects of temperature and
  modifications to the rayleigh-brillouin spectrum, Journal of Physics:
  Conference Series 1239 (2019) 012003.
\newblock \href {http://dx.doi.org/10.1088/1742-6596/1239/1/012003}
  {\path{doi:10.1088/1742-6596/1239/1/012003}}.

\bibitem{perciante-pa-2019}
A.~Garc{\'{\i}}a-Perciante, A.~R. M{\'{e}}ndez, Dissipative properties of
  relativistic two-dimensional gases, Physica A: Statistical Mechanics and its
  Applications 530 (2019) 121559.
\newblock \href {http://dx.doi.org/https://doi.org/10.1016/j.physa.2019.121559}
  {\path{doi:https://doi.org/10.1016/j.physa.2019.121559}}.

\bibitem{heinz-prc-2006}
U.~Heinz, H.~Song, A.~K. Chaudhuri, Dissipative hydrodynamics for viscous
  relativistic fluids, Phys. Rev. C 73 (2006) 034904.
\newblock \href {http://dx.doi.org/10.1103/PhysRevC.73.034904}
  {\path{doi:10.1103/PhysRevC.73.034904}}.

\bibitem{romatschke-prl-2007}
P.~Romatschke, U.~Romatschke, Viscosity information from relativistic nuclear
  collisions: How perfect is the fluid observed at rhic?, Phys. Rev. Lett. 99
  (2007) 172301.
\newblock \href {http://dx.doi.org/10.1103/PhysRevLett.99.172301}
  {\path{doi:10.1103/PhysRevLett.99.172301}}.

\bibitem{song-prc-2008}
H.~Song, U.~Heinz, Causal viscous hydrodynamics in 2 + 1 dimensions for
  relativistic heavy-ion collisions, Phys. Rev. C 77 (2008) 064901.
\newblock \href {http://dx.doi.org/10.1103/PhysRevC.77.064901}
  {\path{doi:10.1103/PhysRevC.77.064901}}.

\bibitem{song-prc-2008b}
H.~Song, U.~Heinz, Multiplicity scaling in ideal and viscous hydrodynamics,
  Phys. Rev. C 78 (2008) 024902.
\newblock \href {http://dx.doi.org/10.1103/PhysRevC.78.024902}
  {\path{doi:10.1103/PhysRevC.78.024902}}.

\bibitem{molnar-jop-2008}
D.~Molnar, P.~Huovinen, Dissipative effects from transport and viscous
  hydrodynamics, Journal of Physics G: Nuclear and Particle Physics 35 (2008)
  104125.
\newblock \href {http://dx.doi.org/10.1088/0954-3899/35/10/104125}
  {\path{doi:10.1088/0954-3899/35/10/104125}}.

\bibitem{denicol-prc-2009}
G.~S. Denicol, T.~Kodama, T.~Koide, P.~Mota, Effect of bulk viscosity on
  elliptic flow near the qcd phase transition, Phys. Rev. C 80 (2009) 064901.
\newblock \href {http://dx.doi.org/10.1103/PhysRevC.80.064901}
  {\path{doi:10.1103/PhysRevC.80.064901}}.

\bibitem{bozek-prc-2010}
P.~Bo\ifmmode~\dot{z}\else \.{z}\fi{}ek, Bulk and shear viscosities of matter
  created in relativistic heavy-ion collisions, Phys. Rev. C 81 (2010) 034909.
\newblock \href {http://dx.doi.org/10.1103/PhysRevC.81.034909}
  {\path{doi:10.1103/PhysRevC.81.034909}}.

\bibitem{dobado-prd-2012}
A.~Dobado, J.~M. Torres-Rincon, Bulk viscosity and the phase transition of the
  linear sigma model, Phys. Rev. D 86 (2012) 074021.
\newblock \href {http://dx.doi.org/10.1103/PhysRevD.86.074021}
  {\path{doi:10.1103/PhysRevD.86.074021}}.

\bibitem{hostler-prc-2013}
J.~Noronha-Hostler, G.~S. Denicol, J.~Noronha, R.~P.~G. Andrade, F.~Grassi,
  Bulk viscosity effects in event-by-event relativistic hydrodynamics, Phys.
  Rev. C 88 (2013) 044916.
\newblock \href {http://dx.doi.org/10.1103/PhysRevC.88.044916}
  {\path{doi:10.1103/PhysRevC.88.044916}}.

\bibitem{kadam-npa-2015}
G.~P. Kadam, H.~Mishra, Bulk and shear viscosities of hot and dense hadron gas,
  Nuclear Physics A 934 (2015) 133 -- 147.
\newblock \href {http://dx.doi.org/10.1016/j.nuclphysa.2014.12.004}
  {\path{doi:10.1016/j.nuclphysa.2014.12.004}}.

\bibitem{karsch-prb-2008}
F.~Karsch, D.~Kharzeev, K.~Tuchin, Universal properties of bulk viscosity near
  the qcd phase transition, Physics Letters B 663~(3) (2008) 217--221.
\newblock \href {http://dx.doi.org/10.1016/j.physletb.2008.01.08}
  {\path{doi:10.1016/j.physletb.2008.01.08}}.

\bibitem{harvey-prl-2008}
H.~B. Meyer, Calculation of the bulk viscosity in su(3) gluodynamics, Phys.
  Rev. Lett. 100 (2008) 162001.
\newblock \href {http://dx.doi.org/10.1103/PhysRevLett.100.162001}
  {\path{doi:10.1103/PhysRevLett.100.162001}}.

\bibitem{dou-aa-2011}
X.~Dou, X.~Meng, Bulk viscous cosmology: Unified dark matter, Advances in
  Astronomy 2011.
\newblock \href {http://dx.doi.org/10.1155/2011/829340}
  {\path{doi:10.1155/2011/829340}}.

\bibitem{gagnon-jcap-2011}
J.~Gagnon, J.~Lesgourgues, Dark goo: bulk viscosity as an alternative to dark
  energy, Journal of Cosmology and Astroparticle Physics 2011~(09) (2011)
  026--026.
\newblock \href {http://dx.doi.org/10.1088/1475-7516/2011/09/026}
  {\path{doi:10.1088/1475-7516/2011/09/026}}.

\bibitem{velten-prd-2013}
H.~Velten, J.~Wang, X.~Meng, Phantom dark energy as an effect of bulk
  viscosity, Phys. Rev. D 88 (2013) 123504.
\newblock \href {http://dx.doi.org/10.1103/PhysRevD.88.123504}
  {\path{doi:10.1103/PhysRevD.88.123504}}.

\bibitem{velten-ijgmmp-2014}
H.~Velten, Viscous cold dark matter in agreement with observations,
  International Journal of Geometric Methods in Modern Physics 11~(02) (2014)
  1460013.
\newblock \href {http://dx.doi.org/10.1142/S0219887814600135}
  {\path{doi:10.1142/S0219887814600135}}.

\bibitem{atreya-jcap-2018}
A.~Atreya, J.~Bhatt, A.~Mishra, Viscous self interacting dark matter and cosmic
  acceleration, Journal of Cosmology and Astroparticle Physics 2018~(02) (2018)
  024--024.
\newblock \href {http://dx.doi.org/10.1088/1475-7516/2018/02/024}
  {\path{doi:10.1088/1475-7516/2018/02/024}}.

\bibitem{gabbana-arxiv-2019}
A.~Gabbana, D.~Simeoni, S.~Succi, R.~Tripiccione, Relativistic lattice
  boltzmann methods: Theory and applications\href
  {http://arxiv.org/abs/1909.04502} {\path{arXiv:1909.04502}}.

\bibitem{anderson-witting-ph-1974a}
J.~Anderson, H.~Witting, A relativistic relaxation-time model for the boltzmann
  equation, Physica 74~(3) (1974) 466 -- 488.
\newblock \href {http://dx.doi.org/10.1016/0031-8914(74)90355-3}
  {\path{doi:10.1016/0031-8914(74)90355-3}}.

\bibitem{anderson-witting-ph-1974b}
J.~Anderson, H.~Witting, Relativistic quantum transport coefficients, Physica
  74~(3) (1974) 489 -- 495.
\newblock \href {http://dx.doi.org/10.1016/0031-8914(74)90356-5}
  {\path{doi:10.1016/0031-8914(74)90356-5}}.

\bibitem{cercignani-book-2002}
C.~Cercignani, G.~M. Kremer, The Relativistic Boltzmann Equation: Theory and
  Applications, Birkhäuser Basel, 2002.
\newblock \href {http://dx.doi.org/10.1007/978-3-0348-8165-4}
  {\path{doi:10.1007/978-3-0348-8165-4}}.

\bibitem{chapman-book-1970}
S.~Chapman, T.~G. Cowling, The Mathematical Theory of Non-Uniform Gases, 3rd
  ed, Cambridge University Press, 197.
\newblock \href {http://dx.doi.org/10.1119/1.1942035}
  {\path{doi:10.1119/1.1942035}}.

\bibitem{grad-cpam-1949}
H.~Grad, On the kinetic theory of rarefied gases, Communications on Pure and
  Applied Mathematics 2~(4) (1949) 331--407.
\newblock \href {http://dx.doi.org/10.1002/cpa.3160020403}
  {\path{doi:10.1002/cpa.3160020403}}.

\bibitem{plumari-prc-2012}
S.~Plumari, A.~Puglisi, F.~Scardina, V.~Greco, Shear viscosity of a strongly
  interacting system: Green-kubo correlator versus chapman-enskog and
  relaxation-time approximations, Phys. Rev. C 86 (2012) 054902.
\newblock \href {http://dx.doi.org/10.1103/PhysRevC.86.054902}
  {\path{doi:10.1103/PhysRevC.86.054902}}.

\bibitem{coelho-cf-2018}
R.~C. Coelho, M.~Mendoza, M.~M. Doria, H.~J. Herrmann, Fully dissipative
  relativistic lattice boltzmann method in two dimensions, Computers \& Fluids
  172 (2018) 318 -- 331.
\newblock \href {http://dx.doi.org/10.1016/j.compfluid.2018.04.023}
  {\path{doi:10.1016/j.compfluid.2018.04.023}}.

\bibitem{gabbana-pre-2017}
A.~Gabbana, M.~Mendoza, S.~Succi, R.~Tripiccione, Towards a unified lattice
  kinetic scheme for relativistic hydrodynamics, Phys. Rev. E 95 (2017) 053304.
\newblock \href {http://dx.doi.org/10.1103/PhysRevE.95.053304}
  {\path{doi:10.1103/PhysRevE.95.053304}}.

\bibitem{denicol-arxiv-2016}
G.~S. Denicol, J.~Noronha, Divergence of the chapman-enskog expansion in
  relativistic kinetic theory~(August).
\newblock \href {http://arxiv.org/abs/1608.07869} {\path{arXiv:1608.07869}}.

\bibitem{noronha-npa-2017}
J.~Noronha, G.~S. Denicol, The onset of fluid-dynamical behavior in
  relativistic kinetic theory, Nuclear Physics A 967 (2017) 417 -- 420.
\newblock \href {http://dx.doi.org/10.1016/j.nuclphysa.2017.05.041}
  {\path{doi:10.1016/j.nuclphysa.2017.05.041}}.

\bibitem{florkowski-prc-2018}
W.~Florkowski, B.~Friman, A.~Jaiswal, E.~Speranza, Relativistic fluid dynamics
  with spin, Phys. Rev. C 97 (2018) 041901.
\newblock \href {http://dx.doi.org/10.1103/PhysRevC.97.041901}
  {\path{doi:10.1103/PhysRevC.97.041901}}.

\bibitem{florkowski-prd-2018}
W.~Florkowski, B.~Friman, A.~Jaiswal, R.~Ryblewski, E.~Speranza, Spin-dependent
  distribution functions for relativistic hydrodynamics of spin-$\frac{1}{2}$
  particles, Phys. Rev. D 97 (2018) 116017.
\newblock \href {http://dx.doi.org/10.1103/PhysRevD.97.116017}
  {\path{doi:10.1103/PhysRevD.97.116017}}.

\bibitem{florkowski-arxiv-2019}
W.~Florkowski, B.~Friman, A.~Jaiswal, R.~Ryblewski, E.~Speranza, Relativistic
  fluid dynamics of spin-polarized systems of particles, arXiv
  preprint~(January).
\newblock \href {http://arxiv.org/abs/1901.00352} {\path{arXiv:1901.00352}}.

\bibitem{niemi-prl-2011}
H.~Niemi, G.~S. Denicol, P.~Huovinen, E.~Moln\'ar, D.~H. Rischke, Influence of
  shear viscosity of quark-gluon plasma on elliptic flow in ultrarelativistic
  heavy-ion collisions, Phys. Rev. Lett. 106 (2011) 212302.
\newblock \href {http://dx.doi.org/10.1103/PhysRevLett.106.212302}
  {\path{doi:10.1103/PhysRevLett.106.212302}}.

\end{thebibliography}


\end{document}